\documentclass[twocolumn,amsmath,amssymb,aps,prl,floatfix,10pt,showpacs]{revtex4-1}
\usepackage[utf8]{inputenc}
\usepackage{graphicx}
\usepackage{dcolumn}
\usepackage{bm}
\usepackage{booktabs}
\usepackage{mathrsfs}
\usepackage{braket}
\usepackage{subfigure}
\usepackage{hyperref}
\usepackage[usenames,dvipsnames]{xcolor}
\usepackage{pdfpages}

\hypersetup{pdfauthor={Benjamin Siegert, Andrea Donarini, and Milena Grifoni},pdftitle={Non-equilibrium spin-crossover in copper phthalocyanine},pdfproducer={unknown}}
 
\allowdisplaybreaks

\newcommand{\erz}[2]{\hat{\operatorname{#1}}_{#2}^{\dag}}			
\newcommand{\ver}[2]{\hat{\operatorname{#1}}_{#2}^{\phantom{\dag}}}		
\newcommand{\rr}{\mathbf{r}}							
\newcommand{\kk}{\mathbf{k}}							
\newcommand{\HH}[1]{\hat{\operatorname{H}}_{\mathrm{#1}}}		  	
\newcommand{\eps}[1]{\epsilon_{#1}}						

\begin{document}

\title{Non-equilibrium spin-crossover in copper phthalocyanine}
\author{Benjamin Siegert}
\email{benjamin.siegert@ur.de}

\author{Andrea Donarini}
\author{Milena Grifoni}
\affiliation{Institut für Theoretische Physik, Universit\"at Regensburg, D-93040 Regensburg, Germany}

\date{\today}

%
%

\begin{abstract}
We demonstrate the tip induced control of the spin state of copper phthalocyanine (CuPc)
on an insulator coated substrate. Accounting for electronic correlations, we find that,
under the condition of energetic proximity of neutral excited states
to the anionic groundstate, the system can undergo a population inversion
towards these excited states. The resulting state of the system is accompanied
by a change in the total spin quantum number. Experimental signatures of the crossover
are the appearance of additional nodal planes in the topographical STM images
as well as a strong suppression of the current near the center of the molecule.
The robustness of the effect against moderate charge conserving relaxation processes
has also been tested.
\end{abstract}

\maketitle

%
%

\emph{Introduction} -
Research on single molecule junctionshas witnessed in recent years a broadening interdisciplinary interest~\cite{Aradhya2013}.
For example, spin dependent transport~\cite{Bogani2008,Sanvito2011} or nuclear spin resonance~\cite{Thiele2014}
have been investigated.
In this emergent field of molecular spintronics, spin-crossover metalorganic compounds (SCOs) play a
prominent role~\cite{Zyazin2010,Osorio2010,Meded2011,sco_fepciv,sco_fepcv}.
These molecules undergo a transition between metastable spin states under the influence of
external stimuli~\cite{sco_guetlich}. The many-body exchange interaction of the $d$-electrons
on the metal center, in combination with the crystal field generated by the surrounding
ligand, determines their spin state.
In three-terminal devices, the change of charge state tuned by the gate electrode
has been shown to govern the associated spin state~\cite{Zyazin2010,Osorio2010,Meded2011}.
Recently, SCOs have been in the focus of STM experiments~\cite{sco_fepciv,sco_fepcv,Gruber2014}.
More generally, the role of many-body effects in STM single molecule junctions is receiving increasing attention,
both theoretically~\cite{benzene_sandra,Donarini2012,Toroz2011,Toroz2013,nature_STM} and experimentally~\cite{repp_charge_state,nature_STM,sco_fepciv,sco_fepcv}.

%
%
%
\begin{figure}[ht!]
\includegraphics[width=\columnwidth]{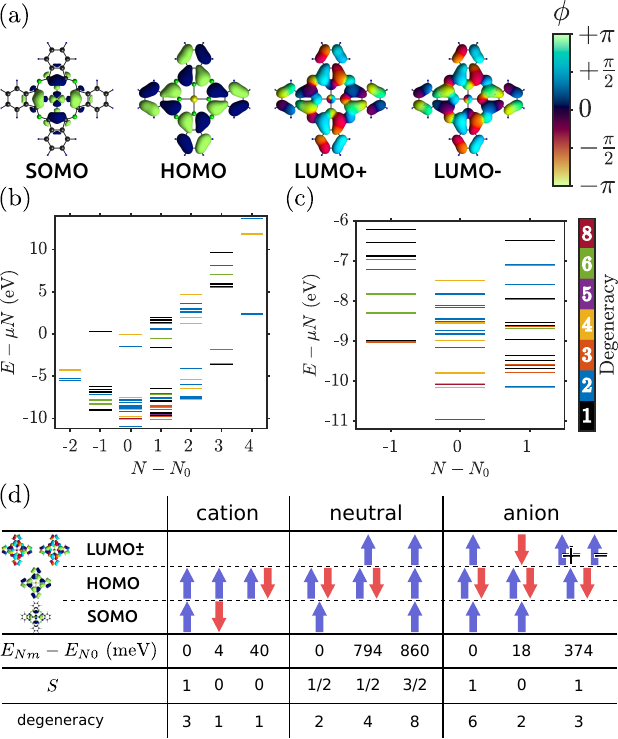}
\caption{(Color online) (a) Frontier orbitals used for the many-body calculation, in their complex representation. The color code shows the phase of the wavefunctions.
	(b), (c) Full and low-energy cutout, respectively, of the many-body spectrum of CuPc at chemical potential $\mu=-4.65~$eV.
	(d) Scheme of the lowest-lying many body states. \label{fig_1} }
\end{figure}
%
In this Letter we demonstrate the appearance of a non-equilibrium high-spin state in
CuPc on an insulating substrate caused by population inversion, and show
experimentally observable fingerprints of this effect.
We illustrate that, for a given substrate work function,
it is possible to control the effective ground state of the molecule by varying the tip position or the bias voltage across the junction.
The only requirements for this genuine many-body effect are an asymmetry between
tip and substrate tunneling rates, which is naturally inherent to STM setups,
and an energetic proximity of an excited neutral state of the molecule
to its anionic ground state. As discussed below, the experimental set-up is similar to that
of Ref.~\cite{repp_charge_state}, but with a slightly larger workfunction for the substrate.
Control over the workfunction can be achieved by choosing different materials or crystallographic
orientation for the substrate, with effects analogous to a discrete gating of the molecule. Several approaches
to gate an STM junction have been also very recently investigated~\cite{Fernandez-Torrente2012,Bouvron2014,Martinez-Blanco2015}.
%
%

\emph{Many-body Hamiltonian and spectrum of CuPc - }
To properly describe the many-body electronic structure of CuPc is by itself a nontrivial task,
since the relatively large size of the molecule makes it impossible to diagonalize
exactly a many-body Hamiltonian written in a local, atomic basis
as done for smaller molecules~\cite{wegewijs_schoeller,benzene_dana_georg,Kaasbjerg2011}.
STM transport experiments on single molecules, however, are restricted
to an energy window involving only the low-lying states of the molecule in its neutral,
cationic and anionic configuration, with the equilibrium configuration at zero bias
set by the workfunction $\phi_0$ of the substrate~\cite{repp_charge_state}.
This allows one to use a restricted basis of frontier orbitals to construct the many-body Hamiltonian~\cite{ryndyk}.
For example, for a copper substrate as in~\cite{repp_charge_state} is $\phi_0=4.65~$eV,
and CuPc in equilibrium is in its neutral ground state.
Thus, in the following we only retain four frontier orbitals of CuPc,
the SOMO ($S$), the HOMO ($H$) and the two degenerate LUMO ($L^\pm$) orbitals, see Fig. 1(a).
In equilibrium, the molecule contains $N_0=3$ frontier electrons.
In this basis, \emph{all} matrix elements of the Coulomb interaction are retained.
Hence, besides Hubbard-like density-density interaction terms,
our model also includes exchange and pair hopping terms, which
ultimately are important for the structure and spin configuration of the molecular excited states.

The Hamiltonian of CuPc in the basis of the four single particle frontier orbitals reads
\begin{align}\label{Eq:H_mol}
	\HH{mol}  =  \sum_{i} \tilde{\epsilon}_{i}\, \hat n_{i} + \frac{1}{2} \sum_{ijkl}\sum_{\sigma\sigma'} V_{ijkl}\,\erz{d}{i\sigma}\erz{d}{k\sigma'}\ver{d}{l\sigma'}\ver{d}{j\sigma},
 \end{align}
where $i=S,H,L\pm$  and $\sigma$ is the spin degree of freedom. The energies $\tilde{\epsilon}_{i}=\eps{i}+\Delta_i$ contain the single particle molecular energies $\eps{i}$ obtained from
diagonalizing the single particle Hamiltonian $\HH{0}$ of CuPc, $\eps{S}=-12.0~$eV, $\eps{H}=-11.7~$eV and $\eps{L\pm}=-10.7~$eV. The parameters $\Delta_i$
account for crystal field corrections and the ionic background of the molecule, since the atomic onsite energies in $\HH{0}$ come from Hartree-Fock calculations for isolated atoms ~\cite{mann_hf}. The $\Delta_i$ are free parameters of the theory.
Isolated CuPc has $D_{4h}$ symmetry; the four molecular orbitals $\ket{i\sigma}$ that make up the basis of Eq.~($\ref{Eq:H_mol}$) transform like its $b_{1g}$ ($S$), $a_{1u}$ ($H$) and $e_u$ ($L^\pm$) representations.
As a consequence, they acquire distinct phases $\phi_i$ when rotated by 90 degrees around the main symmetry axis of the molecule, as illustrated in Fig.~\ref{fig_1}(a).
This yields an easy rule to determine the nonvanishing Coulomb matrix elements $V_{ijkl}$ in  Eq. (\ref{Eq:H_mol}): $V_{ijkl}\neq0$ if $\phi_i-\phi_j+\phi_k-\phi_l = 0~\mathrm{mod}~2\pi$,
i.e., nonvanishing contributions are only possible if the phases of the corresponding molecular orbitals add up
to multiples of $2\pi$. These considerations remain true in the presence of a homogenous substrate which reduces the symmetry to $C_{4v}$.
%
For a detailed discussion concerning the parametrization of Eq.~(\ref{Eq:H_mol}) we refer to the supplemental material~\cite{suppl}.
Exact numerical diagonalization of $\HH{mol}$ finally yields the many body eigenenergies $E_{Nm}$ and eigenstates $\ket{Nm}$
of the molecule, labelled after particle number $N$ and state index $m$.

Since the molecule is in contact with the substrate and is able to exchange electrons, it is necessary to consider a grandcanonical ensemble $\HH{mol}-\mu\hat N$,
where $\mu$ is the chemical potential of the substrate which is given by its negative workfunction, $\mu=-\phi_0$.
Moreover, the presence of the leads renormalizes the Hamiltonian $\HH{0}$ due to image charges effects~\cite{Kaasbjerg2011,Perrin2013}.
We model these effects with an effective Hamiltonian $\HH{\rm mol-env} = -\delta_{\rm ic} (\hat{N}-N_0)^2$, with $\hat{N}$ the particle number operator
on the system and $\delta_{\rm ic}$ obtained from electrostatic considerations, see supplemental material. To fit our spectrum to the experiment of Swart et al.~\cite{repp_charge_state},
which was taken on a copper substrate Cu[100] ($\phi_0=4.65~$eV) on a trilayer of NaCl, we used a constant shift $\Delta_i=\Delta=1.83~$eV, a dielectric constant  $\eps{\rm mol}=2.2$
in the evaluation of the matrix elements $V_{ijkl}$, and an image-charge renormalization $\delta_{\rm ic} = 0.32~$eV.

Figures~\ref{fig_1}(b), (c) show the cationic, neutral and anionic subblocks of the many particle spectrum and their degeneracies.
The neutral groundstate has a doublet structure (with total spin $S=\frac{1}{2}$) coming from the doubly filled HOMO and the unpaired spin in the SOMO,
whereas the cationic and anionic
groundstates have triplet structures ($S=1$). The former has a singly filled HOMO,
the latter a singly filled LUMO orbital which form spin triplets (and singlets, $S=0$, for the first excited states) with the singly filled SOMO.
Finally, the orbital degeneracy of the LUMO makes up for an additional twofold multiplicity of the anionic ground and first excited states.
The first excited state of the neutral molecule is found to be also a doublet ($S=\frac{1}{2}$) with additional twofold orbital degeneracy.
Finally, the second excited state
shows a spin quadruplet structure ($S=\frac{3}{2}$) together with twofold orbital degeneracy.
A schematic depiction of these states is shown in Fig.~\ref{fig_1}(d). As the actual states are linear combinations of several Slater determinants, only dominant contributions are shown.

\emph{Transport dynamics and spin-crossover - }
The full system is characterized by the Hamiltonian $\HH{} = \HH{mol} + \HH{mol-env} + \HH{S} + \HH{T} + \HH{tun}$, where
$\HH{S}$ and $\HH{T}$ are describing noninteracting electronic reservoirs for substrate (S) and tip (T).
The tunneling Hamiltonian is $\HH{tun} = \sum_{\eta\kk i\sigma} t^\eta_{\kk i}\, \erz{c}{\eta\kk\sigma}\ver{d}{i\sigma} +\mathrm{h.c.}$,
where $\erz{c}{\eta\kk\sigma}$ creates an electron in lead $\eta$ with spin $\sigma$ and momentum $\kk$. The tunneling matrix elements $t^\eta_{\kk i}$ are obtained analogously to Ref.~\cite{benzene_sandra}.
The dynamics is calculated via the Generalized Master Equation for the reduced density operator $\rho_\mathrm{red}=\operatorname{Tr}_{\mathrm{S,T}}\left(\rho\right)$, see Refs.~\cite{benzene_dana_georg,benzene_sandra}.
In particular, we are interested in the state $\rho_\mathrm{red}^\infty$ solving the stationary equation $\mathcal{L}[\rho_\mathrm{red}]=0$, where $\mathcal{L}$ is the Liouvillian superoperator.
%
%
\begin{figure}
\includegraphics[width=\columnwidth]{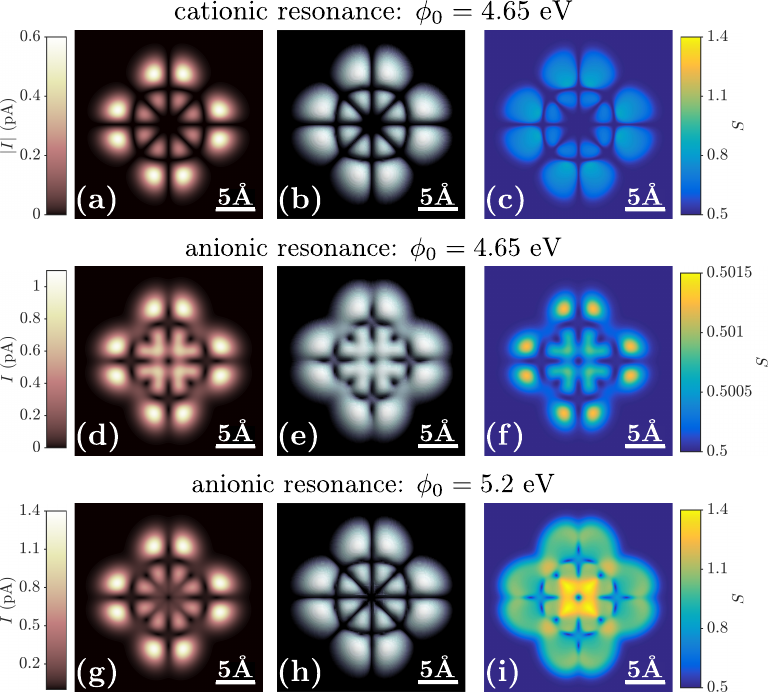}
\caption{(Color online) Constant height current maps (a,d,g), constant current maps (b,e,h) and maps of the system's total spin $S$ (c,f,i).
Constant height and spin maps are taken at a tip-molecule distance of $5~$\AA, constant current maps at currents $I = 0.5,\, 0.75,\,1.0~$pA for panels (c), (f), and (i), respectively. \label{fig_2} }
\end{figure}

In analogy to Ref.~\cite{koch_oppen}, we included a phenomenological relaxation term $\mathcal{L}_{\mathrm{rel}}$ in the Liouvillian \footnote{Differently from Ref.~\cite{koch_oppen}
we have included in \eqref{L_rel} also the coherences. $\mathcal{L}_{\rm rel}$ accounts thus also for dephasing. For simplicity, we assume the same phenomenological rate for dephasing and dissipation.}:
\begin{align}\label{L_rel}
  \mathcal{L}_\mathrm{rel}\left[\rho\right] = -\frac{1}{\tau} \left( \rho  - \sum_{Nm}\rho^{\mathrm{th},N}_{mm}\ket{Nm}\bra{Nm}\sum_n \rho^N_{nn} \right).
\end{align}
It is proportional to the deviation of the reduced density matrix from the thermal one $\rho^\mathrm{th}$, which is given by the
Boltzmann distribution $\rho^{\mathrm{th},N}_{mm}\sim\exp{\left(-\frac{E_{Nm}}{k_BT}\right)}$ with $\sum_m\rho^{\mathrm{th},N}_{mm}=1$.
Since $\mathcal{L}_\mathrm{rel}$ describes relaxation processes which conserve the particle number on the molecule, it does not contribute directly to the current.
The relaxation factor $\frac{1}{\tau}$ is taken of the same order of magnitude of the tip tunneling rate.
The stationary current through the system is evaluated as
\begin{align}\label{Eq_current}
\braket{\hat I_\mathrm{S}+\hat I_\mathrm{T}}=\frac{\mathrm{d}}{\mathrm{d}t}\braket{\hat N}=\operatorname{Tr}_\mathrm{mol}\left(\hat N\mathcal{L}[\rho_\mathrm{red}^\infty]\right)\equiv0,
\end{align}
The Liouvillian $\mathcal{L}=\mathcal{L}_\mathrm{rel}+\sum_\eta\mathcal{L}_\eta$ decomposes into the relaxation term and sub-Liouvillians for each lead.
Sorting of the occuring terms in Eq.~(\ref{Eq_current}) after substrate and tip contributions yields the current operator of the respective lead $\eta$ as $\hat I_\eta=\hat N\mathcal{L_\eta}$.

%
%
\begin{figure}
\includegraphics[width=\columnwidth]{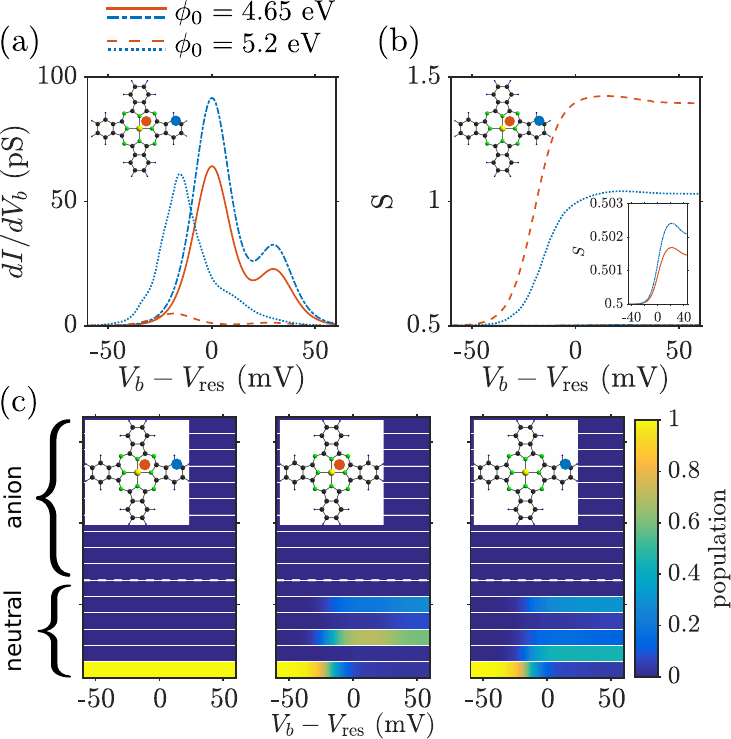}
\caption{(Color online) (a) Differential conductance and (b) total spin curves taken at different tip positions and workfunctions around
the bias $V_\mathrm{res}(\phi_0)$ of the anionic resonance. The inset in (b) shows the change of the spin for the standard case in magnification. %
(c) Populations of the density matrix around $V_\mathrm{res}(\phi_0)$. Left panel: standard case, $\phi_0=4.65~$eV. %
Middle (right) panel: anomalous case, $\phi_0=5.2~$eV, with tip near the center (outer on the ligand). %
 \label{fig_3} }
\end{figure}
Results of our transport calculations are presented in Fig.~\ref{fig_2}.
In panels (a,d,g) we show constant height current maps, constant current STM images in (b,e,h) and in (c,f,i) maps of
the expectation value of the total spin of the molecule depending on the  tip position,
$S_{\rr_\mathrm{T}} = \sqrt{\braket{\hat S^2}_{\rr_\mathrm{T}} +\frac{1}{4} }-\frac{1}{2}$
where $\braket{\hat S^2}_{\rr_\mathrm{T}} = \operatorname{Tr}_\mathrm{mol}\left(\hat S^2\rho_\mathrm{red}^\infty(\rr_\mathrm{T})\right)$.
The constant height and spin maps are each taken at a tip-molecule distance of 5~\AA.
The upper three panels (a,b,c) are for a workfunction of $\phi_0=4.65~$eV and a bias voltage of $V_b=-2.72~$V. At this position the cationic resonance is occuring.
Since the difference between neutral and cationic groundstate is the occupation of the HOMO (see Fig.~\ref{fig_1}(d)), tunneling occurs via this orbital and the current maps (a,b)
resemble its structure. With the same work function $\phi_0=4.65~$eV, the anionic resonance is taking place at positive bias $V_b=0.81~$V, see Fig.~\ref{fig_2}(d,e).
For equivalent reasons as in the former case, tunneling is happening via the LUMO and the spatial dependence of the current resembles the topography of this orbital.
Panels (g,h,i) are recorded, instead, at $\phi_0=5.2~$eV, again at the anionic resonance which is now shifted to $V_b = 1.74~$V
due to the larger workfunction. Panel (g) is puzzling. Despite being an anionic resonance, it closely resembles the HOMO, cf. as panels (a)-(b).
A closer inspection, though, reveals also an alikeness with the LUMO (see panel (d)) but with additional diagonal nodal planes, matching the nodal plane structure of the HOMO.
When observing in panel (h) the constant current map, and comparing it with panels (b) and (e), this statement becomes more evident. This anomalous topography can not be explained by single orbital tunneling.

Panels (c), (f) and (i) reveal the tip-position dependent expectation value of the total spin. At the standard anionic transition, panel (f), the spin remains essentially constant.
At the standard cationic transition, panel (c), a rather homogeneous enhancement of the molecular spin is due to small populations of a large number of excited states,
made accessible by the large resonance bias ($V_{\rm res} = -2.7~$V). The anomalous anionic transition, panel (i), shows, however, the largest variation of the molecular spin,
concentrated at the positions of the anomalous current suppression, compare panels (g) and (d).
To explain the unconventional properties shown in Fig.~\ref{fig_2}, we examine bias traces taken at different tip positions and values of the workfunction.
Figure~\ref{fig_3}(a) shows a shift of the anionic resonant peak in the $\frac{dI}{dV}$ for the anomalous case. The value $V_\mathrm{res}$ at which the peak
is expected is given by
\begin{equation}
 V_\mathrm{res}(\phi_0) = \frac{1}{\alpha_{\rm T}|e|}\left( E_{N_0+1,0} - E_{N_0,0} -\delta_{\rm ic} + \phi_0 \right),
\end{equation}
where $\alpha_{\rm T}$ is the fraction of bias drop between tip and molecule, and $E_{N,0}$ is the energy of the $N$-particle ground state.
The shift of the resonance to lower biases seen in Fig.~\ref{fig_3}(a) suggests the appearance of a population inversion from the neutral ground state to an excited state.
Transitions from the latter to the anionic ground state open in fact at much lower biases.
Also the evolution of the spin of the molecule shown in Fig.~\ref{fig_3}(b) reinforces this proposition.
In the anomalous case, the change of the system from a low to a high spin state, as well as the saturation of the spin, can be clearly seen.
This contrasts the normal anionic transition, where only a marginal change is observable. In Fig.~\ref{fig_3}(c) we show the evolution of the eigenvalues of the stationary
density matrix $\rho_{\rm red}^\infty$, {\it i.e.} the populations of the physical basis~\cite{benzene_dana_georg}, around the anionic resonance $V_\mathrm{res}(\phi_0)$,
depending on workfunction and tip position. In the standard case (left panel of Fig.~\ref{fig_3}(c)), the ground state of the system is always the neutral ground state. For the anomalous case
(middle and right panels of Fig.~\ref{fig_3}) however, the picture changes dramatically, as there is a remarkable depopulation of the neutral ground state in favor of different excited states, depending on the position of the tip.

%
%
\begin{figure}
\includegraphics[width=\columnwidth]{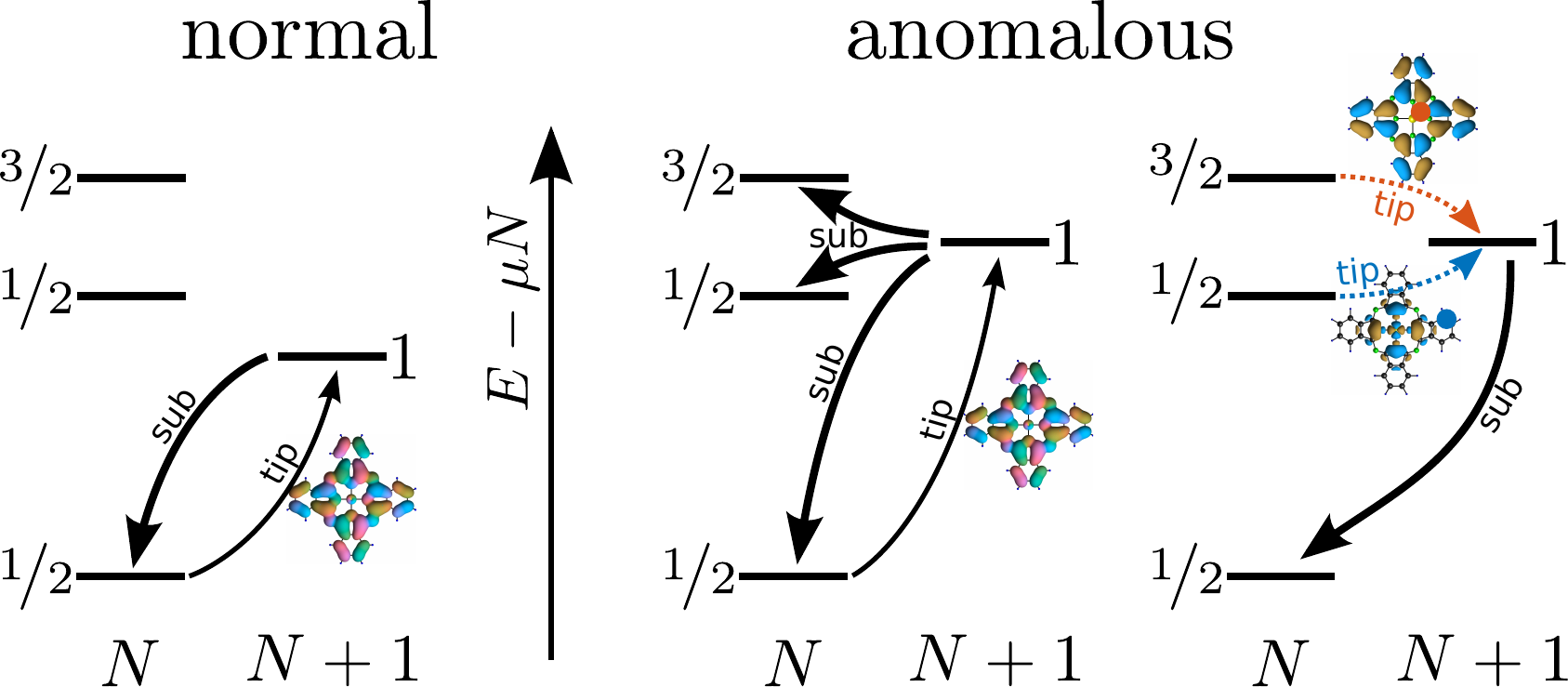}
\caption{(Color online) Simplified sketch of the tunnelling processes at the anionic resonance for the standard ($\phi_0 = 4.65~$eV) and the anomalous ($\phi_0 = 5.2~$eV) case.
In the latter population inversion takes place. The colors of the arrows denote tip positions where the corresponding transition acts as a bottleneck:
Orange (blue) stands for the center (the outer ligand) of CuPc.
 \label{fig_4} }
\end{figure}

We focus now on the mechanism explaining population inversion and associated spin-crossover.
In the standard case, at sufficiently high bias, the transition from the neutral to the anionic groundstate is opening, and tunneling
of an electron into the LUMO brings the molecule into the anionic ground state. By consecutive tunneling to the substrate, the system goes back into its neutral ground state, see Fig.~\ref{fig_4}
for a simple sketch.
Since the tunneling rates to the substrate are much larger than their tip counterparts, the system stays essentially in the neutral ground state with spin $S=\frac{1}{2}$.
Also in the anomalous case an initial tunneling event brings the molecule into the anionic ground state.
However, from there, due to finite temperature and proximity of the many-body eigenenergies, the system has a finite probability to go into a neutral excited state by releasing an electron to the substrate.
The position of the tip and the structure of these excited states themselves then determine the stationary state:
The molecule can only return to its neutral ground state by successive transitions to the anionic ground state via the tip,
and from there to the neutral ground state via the substrate. However, the former process acts as a bottleneck and depends on the tip position.
Leaving the first excited state ($S=\frac{1}{2}$) requires tunneling into the SOMO, while leaving the second excited state ($S=\frac{3}{2}$) would require tunneling into the HOMO.
Additionally, near the center of the molecule the HOMO is vanishing, whereas on the outer ligand part the SOMO has little to no amplitude. Therefore, tunneling into these orbitals at
the respective positions is strongly suppressed and the system ultimately ends up in the corresponding neutral excited states.

%
%
%

\emph{Conclusions - } For an experimentally accessible substrate workfunction of $\phi_0 = 5.2~$eV, we predict the appearance, in proximity to the anionic resonance, of population inversion
between the neutral ground and excited states of CuPc. Depending on the tip position, the molecule is triggered into a low-spin (S=1/2) to high-spin (S=3/2) transition which is
mediated by this population inversion. This inversion is experimentally observable via dramatic changes in the topographical properties
of constant height and constant current STM images, compared to a standard LUMO-mediated anionic transition.
Direct observation of the spin-crossover might be accessible using spin-polarized scanning probe microscopy techniques.~\cite{Pielmeier2013}
The effect is also robust against moderate
charge conserving relaxation processes. The quantitative accuracy of the spectroscopic and topographical results presented in this Letter is limited by the adopted semiempirical model.
The spin-crossover with the associated anomalous topography of the anionic resonance depends, however, on qualitative properties of the many-body spectrum and of the molecular orbitals.
Thus, despite our focus on CuPc, they should be observable also in other molecules with comparable frontier orbital structure.

\begin{acknowledgments}
 The authors thank Thomas Niehaus, Jascha Repp and Dmitry Ryndyk for fruitful discussions.
 Financial support by the Deutsche Forschungsgemeinschaft within the research program SFB 689 is acknowledged.
\end{acknowledgments}

\bibliography{main}

\pagebreak
\widetext



\section*{Supplementary material}

# Non-equilibrium spin-crossover in copper phthalocyanine: Supplemental material

Benjamin Siegert,* Andrea Donarini, and Milena Grifoni
*Institut für Theoretische Physik, Universität Regensburg, D-93040 Regensburg, Germany*
(Dated: July 20, 2015)

## THE SINGLE PARTICLE HAMILTONIAN OF CUPC

The single particle Hamiltonian of CuPc without many-body interaction can be written in second quantization as

$$\hat{H}_0 = \sum_{\substack{\alpha\beta\sigma \\ mn}} h_{\alpha m,\beta n}\, \hat{d}^\dagger_{\alpha m\sigma}\hat{d}_{\beta n\sigma} \tag{1}$$

where $\hat{d}^\dagger_{\alpha m\sigma}$ creates an electron in the atomic orbital $|\alpha m\sigma\rangle$ with quantum numbers $m,\sigma$ centered at atom $\alpha$:

$$\langle \mathbf{r}|\alpha m\sigma\rangle = \phi_{\alpha m\sigma}(\mathbf{r}). \tag{2}$$

The matrix elements are defined as

$$h_{\alpha m,\beta n} := \epsilon_{\alpha m}\delta_{\alpha\beta}\delta_{mn} + b_{\alpha m,\beta n}, \tag{3}$$

where $\epsilon_{\alpha m}$ is the energy of orbital $m$ on atom $\alpha$, and $b_{\alpha m,\beta n}$ is the hopping integral between orbital $m$ on atom $\alpha$ and orbital $n$ on atom $\beta$. The atomic onsite energies are taken from Ref. [1] and the hopping integrals are calculated in nearest-neighbour approximation following the LCAO schemes proposed by Refs. [2, 3], analogously [4] to our earlier work on H$_2$Pc [5]. Geometrical parameters needed for the LCAO calculation like interatomic distances and angles are taken from Ref. [6]. For the ligand we consider the set of all 2s (1s for hydrogen), 2p$_x$ and 2p$_y$ orbitals as the $\sigma$-system, and consequently the set of 2p$_z$ orbitals as the $\pi$-system. On the metal the 3d$_{xy}$, 3d$_{x^2-y^2}$, 3d$_{z^2}$ and 4s orbitals contribute to the $\sigma$-system, while the 3d$_{zx}$ and 3d$_{yz}$ belong to the $\pi$-system. The single particle energies $\epsilon_i$ and molecular orbitals $|i\sigma\rangle$ of the CuPc molecule are obtained by numerical diagonalization of $\hat{H}_0$.

## SETTING UP THE MANY PARTICLE HAMILTONIAN

The full many-body Hamiltonian of a molecule in Born-Oppenheimer approximation in second quantization reads

$$\hat{H}_{\mathrm{mol}} = \hat{H}_0 + \hat{V}_{\mathrm{ee}} = \sum_{\substack{\alpha\beta\sigma \\ mn}} h_{\alpha m,\beta n}\,\hat{d}^\dagger_{\alpha m\sigma}\hat{d}_{\beta n\sigma} + \frac{1}{2}\sum_{\substack{\alpha\beta\gamma\delta \\ mnop}}\sum_{\sigma\sigma'} V^{mnop}_{\alpha\beta\gamma\delta}\,\hat{d}^\dagger_{\alpha m\sigma}\hat{d}^\dagger_{\gamma p\sigma'}\hat{d}_{\delta q\sigma'}\hat{d}_{\beta n\sigma}, \tag{4}$$

where $V^{mnop}_{\alpha\beta\gamma\delta}$ is the matrix element of the Coulomb interaction,

$$V^{mnop}_{\alpha\beta\gamma\delta} = \frac{e^2}{4\pi\epsilon_0}\int \mathrm{d}^3 r_1 \int \mathrm{d}^3 r_2\, \phi^*_{\alpha m\sigma}(\mathbf{r}_1)\,\phi_{\beta n\sigma}(\mathbf{r}_1)\,\frac{1}{|\mathbf{r}_1-\mathbf{r}_2|}\,\phi^*_{\gamma o\sigma'}(\mathbf{r}_2)\,\phi_{\delta p\sigma'}(\mathbf{r}_2). \tag{5}$$

After performing a transformation to the molecular orbital basis, in which $\hat{H}_0$ is diagonal, $\hat{d}^\dagger_{i\sigma} = \sum_{\alpha m} c_{i\alpha m}\,\hat{d}^\dagger_{\alpha m\sigma}$, and using the approximation that the basis $|\alpha m\sigma\rangle$ is orthogonal, i.e. $\langle\alpha m\sigma|\beta n\sigma'\rangle = \delta_{\alpha\beta}\delta_{mn}\delta_{\sigma\sigma'}$, Eq. (4) reads:

$$\hat{H}_{\mathrm{mol}} = \sum_{i\sigma} \epsilon_i\,\hat{n}_{i\sigma} + \frac{1}{2}\sum_{ijkl}\sum_{\sigma\sigma'} V_{ijkl}\,\hat{d}^\dagger_{i\sigma}\hat{d}^\dagger_{k\sigma'}\hat{d}_{l\sigma'}\hat{d}_{j\sigma}, \tag{6}$$

where the matrix element of the Coulomb interaction now reads

$$V_{ijkl} = \frac{e^2}{4\pi\epsilon_0}\int \mathrm{d}^3 r_1 \int \mathrm{d}^3 r_2\, \psi^*_i(\mathbf{r}_1)\psi_j(\mathbf{r}_1)\frac{1}{|\mathbf{r}_1-\mathbf{r}_2|}\psi^*_k(\mathbf{r}_2)\psi_l(\mathbf{r}_2). \tag{7}$$

To cut down the size of our Hilbert space we split the basis into frozen and dynamic states, where $N_f$ frozen states are assumed to be always fully occupied and $N_e$ set to be always empty, whereas we do not make any assumption

about the occupation of the $N_d$ dynamic states. In occupation number representation a general state of the Fock space then looks like

$$|\Psi\rangle \approx |\underbrace{11\ldots 11}_{2N_f}\underbrace{n_{k\uparrow}n_{k\downarrow}\ldots n_{l\uparrow}n_{l\downarrow}}_{2N_d}\underbrace{00\ldots 00}_{2N_e}\rangle. \tag{8}$$

### The different terms of the Coulomb interaction

The Coulomb interaction

$$\sum_{ijkl}\sum_{\sigma\sigma'} V_{ijkl}\,\hat{d}^\dagger_{i\sigma}\hat{d}^\dagger_{k\sigma'}\hat{d}_{l\sigma'}\hat{d}_{j\sigma} \tag{9}$$

can be split into different terms corresponding to the number of different indices in the sum. The first term, where all four indices are the same is:

$$\sum_i V_{iiii}\,n_{i\uparrow}n_{i\downarrow}. \tag{10}$$

Terms with two different indices are:

$$\frac{1}{2}\sum_{[ij]}V_{iijj}\,\hat{n}_i\hat{n}_j + \sum_{[ij]\sigma}\left(V_{iiij}\,\hat{n}_{i\sigma}\hat{d}^\dagger_{i\bar\sigma}\hat{d}^\dagger_{j\bar\sigma}+\text{h.c.}\right) - \frac{1}{2}\sum_{[ij]\sigma}V_{ijji}\left(\hat{n}_{i\sigma}\hat{n}_{j\sigma}-\hat{d}^\dagger_{i\sigma}\hat{d}^\dagger_{j\bar\sigma}\hat{d}_{i\bar\sigma}\hat{d}_{j\sigma}\right) + \frac{1}{2}\sum_{[ij]\sigma}V_{ijij}\,\hat{d}^\dagger_{i\sigma}\hat{d}^\dagger_{i\bar\sigma}\hat{d}_{j\bar\sigma}\hat{d}_{j\sigma}, \tag{11}$$

where we used that $V_{ijkl}=V_{klij}=V^*_{jilk}=V^*_{lkji}$. The symbol $[\ldots]$ under the sum means for example for $[ijk]$: $i\neq j, i\neq k, j\neq k$. Terms with three different indices are:

$$\frac{1}{2}\sum_{[ijk]}\left(V_{ijik}\,\hat{d}^\dagger_{i\sigma}\hat{d}^\dagger_{i\bar\sigma}\hat{d}_{k\bar\sigma}\hat{d}_{j\sigma}+\text{h.c.}\right) + \sum_{[ijk]}\sum_\sigma V_{ijki}\left(\hat{d}^\dagger_{i\sigma}\hat{d}^\dagger_{k\bar\sigma}\hat{d}_{i\bar\sigma}\hat{d}_{j\sigma}-\hat{n}_{i\sigma}\hat{d}^\dagger_{k\sigma}\hat{d}_{j\sigma}\right) + \sum_{[ijk]}\sum_\sigma V_{iijk}\,\hat{n}_i\,\hat{d}^\dagger_{j\sigma}\hat{d}_{k\sigma}. \tag{12}$$

Consequently the last term with four different indices reads

$$\frac{1}{2}\sum_{[ijkl]}\sum_{\sigma\sigma'}V_{ijkl}\,\hat{d}^\dagger_{i\sigma}\hat{d}^\dagger_{k\sigma'}\hat{d}_{l\sigma'}\hat{d}_{j\sigma}. \tag{13}$$

### Symmetry considerations

The molecular orbitals $|i\rangle$ are transforming under symmetry operations of the group of the molecule according to its irreducible representations and are thus acquiring distinct phase factors. Let $R$ be a rotation of 90 degrees around the fourfold molecular axis with $R^{-1}\mathbf{r}=:\tilde{\mathbf{r}}$:

$$\begin{aligned}R|j\rangle &= e^{i\phi_j}|j\rangle,\\ \psi_j(R^{-1}\mathbf{r}) &= \psi_j(\tilde{\mathbf{r}}) = \langle\tilde{\mathbf{r}}|j\rangle = \langle\mathbf{r}|R|j\rangle = e^{i\phi_j}\langle\mathbf{r}|j\rangle = \psi_j(\mathbf{r})\,e^{i\phi_j}.\end{aligned} \tag{14}$$

This yields for $V_{ijkl}$ after a coordinate transformation $\mathbf{r}_i\to\tilde{\mathbf{r}}_i$ under the integral:

$$\begin{aligned}V_{ijkl} &= \frac{e^2}{4\pi\epsilon_0}\int d^3r_1\int d^3r_2\,\psi^*_i(\mathbf{r}_1)\psi_j(\mathbf{r}_1)\frac{1}{|\mathbf{r}_1-\mathbf{r}_2|}\psi^*_k(\mathbf{r}_2)\psi_l(\mathbf{r}_2)\\ &= \frac{e^2}{4\pi\epsilon_0}\int d^3\tilde{r}_1\int d^3\tilde{r}_2\,\psi^*_i(\tilde{\mathbf{r}}_1)\psi_j(\tilde{\mathbf{r}}_1)\frac{1}{|\tilde{\mathbf{r}}_1-\tilde{\mathbf{r}}_2|}\psi^*_k(\tilde{\mathbf{r}}_2)\psi_l(\tilde{\mathbf{r}}_2)\\ &= \frac{e^2}{4\pi\epsilon_0}\int d^3r_1\int d^3r_2\,\psi^*_i(R^{-1}\mathbf{r}_1)\psi_j(R^{-1}\mathbf{r}_1)\frac{1}{|\mathbf{r}_1-\mathbf{r}_2|}\psi^*_k(\tilde{\mathbf{r}}_2)\psi_l(\tilde{\mathbf{r}}_2)\\ &= \frac{e^2}{4\pi\epsilon_0}\int d^3r_1\int d^3r_2\,\psi^*_i(\mathbf{r}_1)e^{-i\phi_i}\psi_j(\mathbf{r}_1)e^{i\phi_j}\frac{1}{|\mathbf{r}_1-\mathbf{r}_2|}\psi^*_k(\mathbf{r}_2)e^{-i\phi_k}\psi_l(\mathbf{r}_2)e^{i\phi_l}\\ &= e^{-i(\phi_i-\phi_j+\phi_k-\phi_l)}\,V_{ijkl}.\end{aligned}$$

| | | | |
|---|---|---|---|
| $U_S$ | 11.352 eV | $J^{\text{ex}}_{HL} = -\tilde{J}^{\text{p}}_{H+-}$ | 548 meV |
| $U_H$ | 1.752 eV | $J^{\text{ex}}_{+-}$ | 258 meV |
| $U_L = U_{+-}$ | 1.808 eV | $J^{\text{p}}_{+-}$ | 168 meV |
| $U_{SH}$ | 1.777 eV | $J^{\text{ex}}_{SL} = -\tilde{J}^{\text{p}}_{S+-}$ | 9 meV |
| $U_{SL}$ | 1.993 eV | $J^{\text{ex}}_{SH} = J^{\text{p}}_{SH}$ | 2 meV |
| $U_{HL}$ | 1.758 eV | | |

TABLE I. Major nonvanishing Coulomb integrals between the SOMO($S$), the HOMO($H$), the LUMO$^+$ ($+$) and the LUMO$^-$ ($-$). All values are calculated numerically using Monte Carlo integration [7] of the real space orbitals depicted in Fig. 1(a) of the main publication and renormalized by a constant $\epsilon_{\text{mol}} = 2.2$.

In order to obtain a true statement the phases have to meet the constraint

$$\phi_i - \phi_j + \phi_k - \phi_l = 0 \mod 2\pi. \tag{15}$$

This condition drastically reduces the number of possible matrix elements of the Coulomb interaction, since the SOMO ($S$), the HOMO ($H$) and the two degenerate LUMO ($\pm$) orbitals gather different phases under rotations of 90 degrees around the main symmetry axis of the CuPc molecule:

$$\phi_S = \pi, \tag{16}$$
$$\phi_H = 0, \tag{17}$$
$$\phi_{L^\pm} = \pm\frac{\pi}{2}. \tag{18}$$

This we will exploit to further simplify Eq. ((6)). Thus, for $V_{ijkl}$ it holds that:

$$V_{ijkl} = e^{-i(\phi_i - \phi_j + \phi_k - \phi_l)} V_{ijkl} \tag{19}$$
$$\Rightarrow V_{ijkl} \neq 0 \quad \text{if} \quad \phi_i - \phi_j + \phi_k - \phi_l = 0 \mod 2\pi, \tag{20}$$

since all phases $\phi_i$ are different; $\phi_i \neq \phi_j$ for $i \neq j$. Putting all together, Eq. ((6)) can be recast in the form

$$\hat{H}_{\text{mol}} = \sum_i (\epsilon_i + \Delta_i)\, \hat{n}_i + \sum_i U_i\, n_{i\uparrow} n_{i\downarrow} + \frac{1}{2}\sum_{[ij]} U_{ij}\, \hat{n}_i \hat{n}_j - \frac{1}{2}\sum_{[ij]}\sum_\sigma J^{\text{ex}}_{ij} \left( \hat{n}_{i\sigma} \hat{n}_{j\sigma} - \hat{d}^\dagger_{i\sigma} \hat{d}^\dagger_{j\bar\sigma} \hat{d}_{i\bar\sigma} \hat{d}_{j\sigma} \right)$$
$$+ \frac{1}{2}\sum_{[ij]}\sum_\sigma J^{\text{p}}_{ij}\, \hat{d}^\dagger_{i\sigma} \hat{d}^\dagger_{i\bar\sigma} \hat{d}_{j\bar\sigma} \hat{d}_{j\sigma} + \frac{1}{2}\sum_{[ijk]}\sum_\sigma \left( \tilde{J}^{\text{p}}_{ijk}\, \hat{d}^\dagger_{i\sigma} \hat{d}^\dagger_{i\bar\sigma} \hat{d}_{k\bar\sigma} \hat{d}_{j\sigma} + \text{h.c.} \right) + \frac{1}{2}\sum_{[ijkl]}\sum_{\sigma\sigma'} V_{ijkl}\, \hat{d}^\dagger_{i\sigma} \hat{d}^\dagger_{k\sigma'} \hat{d}_{l\sigma'} \hat{d}_{j\sigma}, \tag{21}$$

where indices are now running only over orbitals from the dynamic set ($S$, $H$, $L^+$, $L^-$). The abbreviations we introduced in Eq. ((21)) are the orbital Coulomb interaction $U_i = V_{iiii}$, the inter-orbital Coulomb interaction $U_{ij} = V_{iijj}$, the exchange integral $J^{\text{ex}}_{ij} = V_{ijji}$, the ordinary pair hopping term $J^{\text{p}}_{ij} = V_{ijij}$ and the split pair hopping term $\tilde{J}^{\text{p}}_{ijk} = V_{ijik}$. Matrix elements $V_{ijkl}$ of the Coulomb interaction are calculated numerically by Monte Carlo integrating [7] the respective single particle orbitals with the Coulomb interaction $V(\mathbf{r}_1, \mathbf{r}_2) = (4\pi\varepsilon_0 |\mathbf{r}_1 - \mathbf{r}_2|)^{-1}$ and renormalizing the bare values by a constant screening factor $\varepsilon_{\text{mol}} = 2.2$ to account for the screening given by the electrons in the frozen occupied states. As a real space basis for the atomic orbitals $\phi_{\alpha m \sigma}(\mathbf{r})$ Slater-type orbitals [8] were used with effective charges taken from Ref. [9]. Table I lists the major nonvanishing Coulomb matrix elements which were used in this work.

## ELECTROSTATIC CONSIDERATIONS

When depositing the junction on the metallic substrate coated by an insulating film, the parameters of the many-body Hamiltonian introduced in the previous sections are renormalized by image charge effects[10]. Also the presence of the other metallic lead (the tip) modifies the spectrum of the system. An additional source of complexity is introduced by the application of an external bias across the junction which brings the system out of equilibrium. Despite the complexity of a rigorous self-consistent treatment of the electrostatics of a nanojunction, the essential phenomena can be summarized to two: i) the image charges in the nearby metallic leads tend to stabilize an excess of charge on the

molecule, by reducing its addition energy; ii) The polarizability of the molecule implies that the bias voltage between source and drain is not entirely dropping at the contacts between the molecule and the leads. Depending on the geometry of the junction a considerable fraction can also drop across the molecule itself. Considering the weak coupling to the leads and the planar configuration of the CuPc in the STM set up, we have included the previous phenomena in the following minimal model. Firstly, we have introduced the effective Hamiltonian $\hat{H}_{\text{mol}-\text{env}} = -\delta_{\text{ic}}(\hat{N} - N_0)^2$ where $\hat{N}$ counts the number of (frontier) electrons on the molecule, $N_0 = 3$ is the number of (frontier) electron associated to the neutral molecule and $\delta_{\text{ic}}$ is the strength of the image charge renormalization. Secondly, we have assumed the many-body levels obtained from the diagonalization of $\hat{H}_{\text{mol}} + \hat{H}_{\text{mol}-\text{env}}$ to be independent of the applied bias voltage. Moreover, for the bias dependance of the leads chemical potentials, we have used $\mu_{\text{S/T}} = \mu_0 \pm e\alpha_{\text{S/T}}V_b$ but requiring $\alpha_{\text{S}} + \alpha_{\text{T}}\alpha_{\text{M}} = 1$. The values of $\delta_{\text{ic}}$, and $\alpha_{\text{S/T/M}}$ are calculated according to the following electrostatic considerations.

From the addition energy of the neutral molecule $U_0 \equiv E_{N_0+1,0} - 2E_{N_0,0} + E_{N_0-1,0}$ we associate a capacitance to the molecule $C_{\text{M}} = e^2/U_0$. For the tip-molecule and the substrate-molecule capacitances we adopt the parallel plate model and define $C_{\text{T}} = \epsilon_0 A/h$ and $C_{\text{S}} = \epsilon_0\epsilon_r A/d$, where $A = 144$ Å$^2$ is an estimate of the CuPc single molecule surface, $h$ is the tip-molecule distance, $\epsilon_r = 5.9$ is the relative permittivity of NaCl, $d$ is the thickness of the NaCl thin film and $\epsilon_0$ is the vacuum permittivity. Connecting these three capacitances in series, we obtain an estimate of the relative potential drops

$$\alpha_{\text{S/T/M}} = \frac{C_{\text{tot}}}{C_{\text{S/T/M}}} \tag{22}$$

where $C_{\text{tot}}^{-1} = C_{\text{S}}^{-1} + C_{\text{T}}^{-1} + C_{\text{M}}^{-1}$. The relative potantial drop on the molecule $\alpha_{\text{M}}$ for $h = 5$ Å and $d = 8.1$ Å (thickness of a trilayer NaCl) and $U_0 = 2.7$ eV is about a quarter of the applied bias.

The estimate for the image charge parameter $\delta_{\text{ic}}$ proceeds from the same model. First we calculate the electrostatic energy associated to the three capacitors $C_{\text{T}}$, $C_{\text{M}}$ and $C_{\text{S}}$ in series when: i) No external bias is applied but the first and the last plate are grounded and ii) a unit charge is deposited between $C_{\text{T}}$ and $C_{\text{M}}$ ($C_{\text{M}}$ and $C_{\text{S}}$). We call the associated electrostatic energy $E_{\text{up}}$ ($E_{\text{down}}$). From a simple calculation one obtains:

$$E_{\text{up}} = \frac{e^2}{2}\frac{1}{\frac{C_{\text{M}}C_{\text{S}}}{C_{\text{M}}+C_{\text{S}}} + C_{\text{T}}}, \qquad E_{\text{down}} = \frac{e^2}{2}\frac{1}{\frac{C_{\text{M}}C_{\text{T}}}{C_{\text{M}}+C_{\text{T}}} + C_{\text{S}}} \tag{23}$$

Finally, for the image charge parameter we write:

$$-\delta_{\text{ic}} = \frac{E_{\text{up}} + E_{\text{down}}}{2} - U_0/2, \tag{24}$$

*i.e.* the difference between the average electrostatic energy and the average energy needed to charge the isolated molecule. The average of the electrostatic energies gives an estimate of the energy needed to charge the molecule in presence of the leads. If we subtract from it the average addition energy of the isolated molecule $U_0/2$, for which we already account in the many-body Hamiltonian (21), we obtain indeed an estimate of the image charge effects.

Remarkably, using the model that we just exposed, we could fit the absolute spectroscopic position of the anionic and cationic transition and obtain, in accordance with the experiments, standard topographical images for CuPc on Cu[100] and trilayer NaCl [11], as well as CuPc on Cu[111] and bilayer NaCl [12] within essentially the same set of fitting parameters $\epsilon_{\text{mol}}$ and $\Delta$ for the isolated molecule, see Table I.

---

|  | NaCl(3ML) Cu[100] | NaCl(2ML) Cu[111] |
|---|---|---|
| $\phi_0$ (eV) | 4.65 | 5.00 |
| $d$ (Å) | 8.1 | 6.0 |
| $\Delta$ (eV) | 1.83 | 1.74 |
| $\delta_{\text{ic}}$ (eV) | 0.32 | 0.44 |
| $\alpha_{\text{S}}$ | 0.16 | 0.12 |
| $\alpha_{\text{T}}$ | 0.59 | 0.62 |
| $V_{\text{an}}^{\text{exp}}$ (V) | 0.81 | 0.95 |
| $V_{\text{an}}^{\text{th}}$ (V) | 0.81 | 1.01 |
| $V_{\text{cat}}^{\text{exp}}$ (V) | -2.62 | -2.15 |
| $V_{\text{cat}}^{\text{th}}$ (V) | -2.72 | -2.00 |

TABLE II. The table contains i) The substrate tabulated parameters: work function $\phi_0$ and insulating layer thickness $d$ ii) The fitting parameter: crystal field energy shift $\Delta$. The molecular relative permittivity is taken $\epsilon_{\text{mol}} = 2.2$ in both cases. iii) The relative potential drops $\alpha_{\text{S/T}}$ and the image charge renormalization $\delta_{\text{ic}}$ are calculated according to (22) and (24), assuming, in both cases, a tip molecule distance $h = 5$ Å and a molecular surface $A = 144\text{Å}^2$ iv) Experimental and fitted values for the biases corresponding to the anionic and cationic resonances. The experimetal values are extracted from [11] for NaCl(3ML)/Cu[100] and from [12] for NaCl(2ML)/Ci[111].



\end{document}